\documentclass[conference]{IEEEtran}
\usepackage{cite}
\usepackage{amsmath,amssymb,amsfonts}
\usepackage{algorithmic}
\usepackage{graphicx}
\usepackage{textcomp}
\usepackage{xcolor}
\usepackage{hyperref} 
\usepackage{booktabs} 
\usepackage{subfigure} 
\usepackage{balance} 
\usepackage{multicol}
\usepackage{siunitx}
\usepackage{verbatim}
\def\BibTeX{{\rm B\kern-.05em{\sc i\kern-.025em b}\kern-.08em
    T\kern-.1667em\lower.7ex\hbox{E}\kern-.125emX}}
    
\newcommand{\rf}{Random Forest}
\begin{document}

\title{Forest Packing: Fast, Parallel Decision Forests}

\author{\IEEEauthorblockN{James Browne}
\IEEEauthorblockA{\textit{Department of Computer Science} \\
\textit{Johns Hopkins University}\\
Baltimore, USA \\
james.browne@jhu.edu}
\and
\IEEEauthorblockN{Tyler Tomita}
\IEEEauthorblockA{\textit{Department of Biomedical Engineering} \\
\textit{Johns Hopkins University}\\
Baltimore, USA \\
ttomita@jhu.edu}
\and
\IEEEauthorblockN{Disa Mhembere}
\IEEEauthorblockA{\textit{Department of Computer Science} \\
\textit{Johns Hopkins University}\\
Baltimore, USA \\
disa@jhu.edu}
\and
\IEEEauthorblockN{\color{white}{xxxxxxxxxxxxxxx}}
\IEEEauthorblockA{\textit{} \\
\textit{}\\
 \\
}
\and
\IEEEauthorblockN{Randal Burns}
\IEEEauthorblockA{\textit{Department of Computer Science} \\
\textit{Johns Hopkins University}\\
Baltimore, USA \\
randal@jhu.edu}
\and
\IEEEauthorblockN{Joshua T. ~Vogelstein}
\IEEEauthorblockA{\textit{Department of Biomedical Engineering} \\
\textit{Johns Hopkins University}\\
Baltimore, USA \\
jovo@jhu.edu}
}

\maketitle

\begin{abstract}
Machine learning has an emerging critical role in high-performance computing to modulate simulations, extract knowledge from massive data, and replace numerical models with efficient approximations.  Decision forests are a critical tool because they provide insight into model operation that is critical to interpreting learned results.  While decision forests are trivially parallelizable, the traversals of tree data structures incur many random memory accesses and are very slow.

We present memory packing techniques that reorganize learned forests to minimize cache misses during classification.  The resulting layout is hierarchical.  At low levels, we pack the nodes of multiple trees into contiguous memory blocks so that each memory access fetches data for multiple trees.  At higher levels, we use leaf cardinality to identify the most popular paths through a tree and collocate those paths in cache lines.  We extend this layout with out-of-order execution and cache-line prefetching to increase memory throughput.

Together, these optimizations increase the performance of classification in ensembles by a factor of four over an optimized C++ implementation and a factor of 50 over a popular R-language implementation.
\end{abstract}


\section{Introduction}
High-performance computing (HPC) and machine learning (ML) are co-evolving in complex ways.
Machine learning becomes an HPC application when run at scale and ML systems adopt HPC techniques 
to accelerate and parallelize model training and classification. 
The largest deep networks run on hierarchical clusters and use collective communication and GPUs \cite{parameter_server}.
Also in HPC, numerical simulation, data analysis, and visualization integrate machine learning algorithms to improve performance,
reduce data, or manage workflow.  Deep networks classify gravitational waves more accurately and 
orders of magnitude faster than matched filtering in LIGO simulations \cite{ligo}.

Using reinforcement learning to drive parameter exploration has transformed mapping protein-ligand interactions, reducing simulation
time from hours to minutes \cite{protein}.

We note that in 2017, that the top supercomputing conference had a machine learning workshop \cite{SC17} and the top machine learning conference had an HPC workshop \cite{nips17}.

Despite their extreme utility, decision forests have only been lightly used in supercomputing applications.
Decision forests are a broad class of algorithms that includes gradient boosted forests \cite{xgboost} and random forest \cite{rf}.
They are extremely flexible, tend not to overfit data, and can be used for classification, regression, density estimation, 
and manifold learning \cite{criminisi2011decision}.  
One key benefit of decision forests is that most variants are {\em interpretable}---the variables used in classification
can be observed and visualized.   This is important for scientists to explain results and make comprehensible conclusions. 
This differs from deep learning in which the operation of the learning model is opaque. Each node in the network is a non-linear activation of a large number of inputs.  Also, decision forests use minimal computing resources when compared with deep learning. Recent advances have defined the first scalable implementation \cite{xgboost} integrated with HPC schedulers and message-passing distribution \cite{rabit}.

Yet, the poor memory performance of decision forests during classification remains an obstacle to adopting these techniques for HPC.
Decision forests consist of hundreds to thousands of trees and the classification step in a decision forest 
navigates a path from root to leaf in each tree.  Typically mappings of tree nodes to memory store each tree separately and order nodes either according to some traversal discipline or in order of node creation.  In both layouts, classification workloads result in a cache-miss for almost every node traversed during the prediction process.  So, even though decision forests are computationally efficient, 
performing categorical or numerical comparisons at each node, the latency of cache misses makes 
classification throughput unacceptable for many HPC applications.

We optimize the memory layout of decision trees and combine this with asynchronous memory I/O to overcome performance limitations.
Our system takes a trained forest as input and reorganizes the nodes in memory to reduce cache misses.  
For the top levels of a tree, we pack the nodes from multiple trees into cache lines and 
evaluate multiple trees in parallel so that a single cache miss fetches data for multiple tree nodes.  Deeper in the tree, 
we use the cardinality of leaf nodes to identify popular paths and cluster tree nodes into cache lines.  Clustering
minimizes the total number of cache misses, assuming that inputs are selected uniformly at random from the same 
suitably biased distribution as training data.  We further improve performance by overlapping computation with memory access by executing 
tree nodes out of order when data is ready and issuing prefetch instructions for adjacent cache lines that contain nodes
that will be evaluated.  The encoding of trees in the forest also allows for an efficient OpenMP implementation that maximizes L1 cache use by associating groups of trees to specific processors. 

Our optimizations result in a four times speedup over a popular C++ implementation of random forest, Ranger, and a 50 times speedup over a popular R implementation known for robustness, R-RerF.  We evaluate the system on standard, large ML datasets.  Our treatment includes a breakdown of 5 different memory layout optimizations that identifies their individual contributions.  The implementation is freely-available under a permissive open-source license\cite{rerf,rerf2}.  We have also created a container stored on Docker Hub that reproduces all of the experiments and figures in this paper \cite{docker}.  The specifics of the results will vary with processor and memory architecture, but matching the AWS instances we use in our experiments should closely recreate (not reproduce) our results.

\section{Related Work}

Ho introduced the concept of a random decision forest \cite{ho1998random} as an ensemble learning technique comprising randomized decision trees.
Breiman \cite{rf} extended this concept to search only random subspaces at each node, defining the ``random forest'' algorithm in its most used form.
Gradient boosted forests use gradient boosting, an error based weighting of random subspaces, as the weak learner in the ensemble \cite{boosting}.  
Criminisi et al. \cite{criminisi2011decision} provide a detailed review of the family of algorithms and their diverse application to
classification, regression, manifold learning, and semi-supervised learning.

Most implementations of decision trees are available in R and Python packages, including ranger \cite{ranger} and scikit-learn \cite{scikit}.
XGBoost \cite{xgboost}  is the state of the art implementation of gradient boosted forests and the first serious systems effort to
optimize tree ensembles.  XGBoost leverages sparsity in the inputs and shards and compresses data in a manner that access data in cache 
friendly chunks. XGBoost evaluates each tree individually and does not optimize across trees in the forest.  Thus, forest packing could 
be applied to XGBoost with the caveat that forest packing would need to implement compression and sharding.  Finally, XGBoost takes design
advantage of very shallow trees for gradient boosting.  We consider a broader class of shallow and deep trees and optimizations work for 
all decision forest variants.

The concept of forest packing was inspired by Hilbert-packed R-trees \cite{hprt} in which the nodes of an R tree are placed on disk based in an order derived from a space-filling, Hilbert curve.  The recursive nature of space filling curves clusters objects that are close in space near each other in the index.
Similar to forest-packing, this reduces random I/O.  But, the I/O in question is primarily disk I/O, not memory accesses.

In many learning applications, classifiers are trained once and deployed and used repeatedly.  In this scenario, it makes sense to invest in optimizing the trained model prior to deployment. Our work falls into this category.  The Tensor Processing Unit \cite{tpu} was developed with a similar justification.  Google trains models on distributed clusters of GPUs and then deploys models for classification on a custom, low-power, reduced-precision processor to optimize power and cost for classification.

\section{Methods}
\label{sec:methods}
Forest packing is a system that optimizes the memory layout and instruction scheduling of classification in decision forests.  We use classification to 
indicate model evaluation, rather than model training.  In other variants of decision forests, this may be the regression step or manifold learning.
The input to this process is a trained forest that consists of $T$ decision trees.  Each internal node describes a splitting condition on the data and has two 
children.  We assume that each leaf node contains a single class label.  This is typical of random forests \cite{rf,requirements1}, but not true for all variants.
One of the optimizations that we present relies on the single class assumption.

Forest packing reorganizes the trees to minimize cache misses during a parallel and distributed classification.  
It outputs $T/B$ bins in which each bin contains $B$ trees.  Within each bin, we interleave low-level nodes of multiple trees to realize memory parallelism in each cache line access.  We use split cardinality information to store popular paths contiguously for all other nodes.
During evaluation, each bin of trees is assigned to a OpenMP thread (for shared memory) or process (for distributed memory).

The forest packing system includes a runtime system that prefetches data and evaluates tree nodes out-of-order as data is ready, leveraging the 
memory layout to maximize performance.

\subsection{Memory Layout}
\label{sec:MemLayout}
\begin{figure}[t!]
  \centerline{\includegraphics[width=\linewidth]{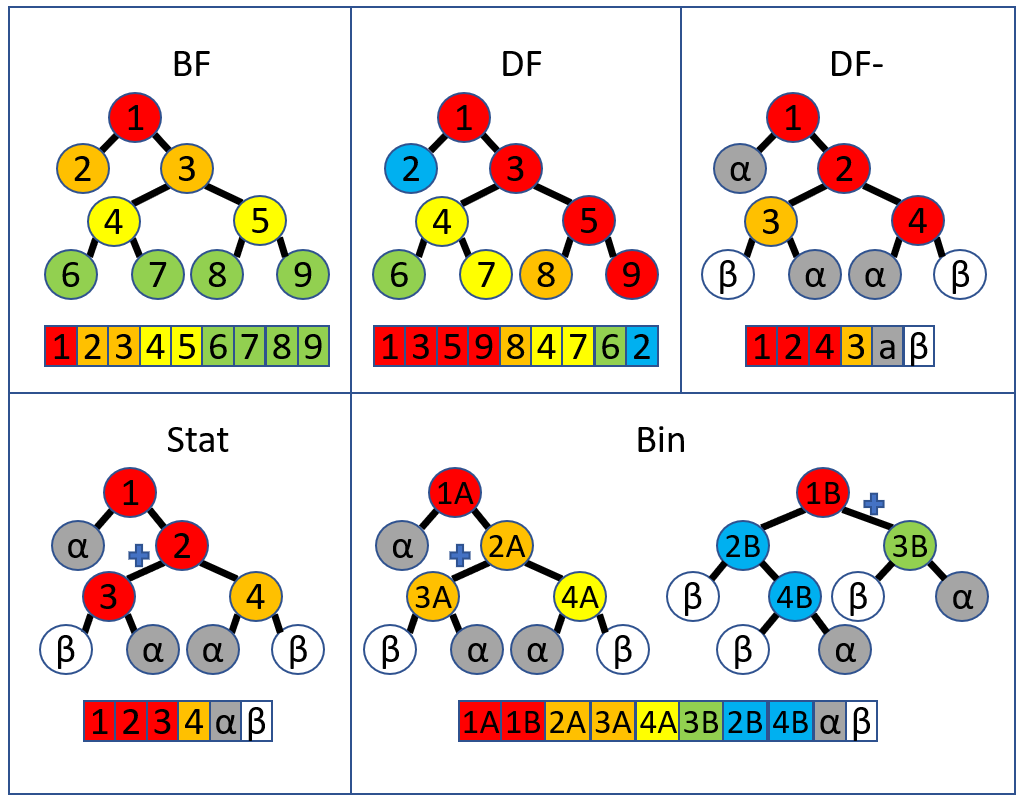}}
  \caption{The representation of trees as arrays of nodes in memory for 
breadth-first \textsf{BF}, 
depth-first \textsf{DF}, 
depth-first with leaf nodes replaced by class nodes \textsf{DF-}, 
statistically ordered depth-first with leaf nodes replaced by class nodes \textsf{Stat}, and 
root nodes interleaved among multiple trees \textsf{Bin}.  Like colors denote a path of processed nodes in the tree.}
  \label{fig:layouts}
\end{figure}

We describe our memory layout as a progression of improvements over the breadth-first layout (\textsf{BF}) used in random forests with each improvement reducing cache misses or encoding parallelism.  Fig.~\ref{fig:layouts} describes this evolution with each panel displaying a tree in which the nodes are numbered breadth-first and the resulting layout itself shown as an array at the bottom of the panel.  Breadth-first layouts are typical in decision forests because they most easily handle leaf nodes at different depths.  Many implementations assume a complete tree of some max depth and then use the statically defined breadth-first layout.  This makes it easy to perform parallel updates within the tree during training.  We include depth-first (\textsf{DF}) as an alternate layout because we will use aspects of both breadth- and depth-first in our ultimate design.

Our first optimization builds on the observation that trees are used differently during training and classification.  During training, an algorithm incrementally populates a tree with data that it routes through the tree splits into leaf nodes in which each leaf node contains a preset minimum number of observations.  Each leaf tracks a class label, determined by the majority observation class, and the cardinality of observations.  For classification, the training data and cardinality does not matter, only the class label. 

We replace the multiple leaf nodes that encode a single class with a single leaf node representing that class, reducing the total size of the tree.  The \textsf{DF-} panel of Fig.~\ref{fig:layouts} shows the resulting layout when building a decision tree for a two-class ($\alpha,\beta$) problem.  \textsf{DF-} indicates depth-first with leaf nodes removed.  In most datasets, the number of distinct classes are many orders of magnitude smaller than the number of nodes in the tree and so we expect the DF- trees to be nearly half the size of the DF trees.

\textsf{DF-} benefits all levels of caches by removing redundant nodes that could otherwise be loaded in addition to requested nodes.  This removal has the dual benefit of allowing a potentially more useful nodes to be loaded (a fact we exploit in the \textsf{Stat} layout) and also greatly reduces cache pollution.
The class nodes that replace leaf nodes are placed at the end of each tree's node array.

Next, we choose which deep paths to encode sequentially in memory based on the frequency of the access.  So, our depth first layout is statistically defined (\textsf{Stat}), rather than according to a traversal order, such as in-order, pre-order, post-order.  We use the leaf cardinalities collected during training to determine the likelihood that any given data point routes to a specific leaf.  These statistics can also be inferred after training using a suitably large set of observations.  We then enumerate depth first paths based on their probability of access.  The \textsf{Stat} panel of Fig.~\ref{fig:layouts} illustrates this process with $+$ indicating a more likely path, leading to the decision to enumerate the path to node $3$ prior to node $4$.  The statistical ordering applies to nodes from the root to the leaf parents.  Again, class nodes that replace leaf nodes are at the end of the array and so are not considered for statistical ordering.

In more detail, the statistical layout, \textsf{Stat}, considers each parent node and its two children.  The child that is accessed most often is placed adjacent to the parent node in memory while the less accessed child is placed later in the node array.  If a parent node has a leaf node child and an internal node child, the child node is always placed adjacent to the parent while the leaf node is shared among all leaf nodes of the same class at the end of the tree's node array.  

To help justify our implementation and provide a method to access alternate encodings we define expected utility of a cache load (\textsf{EU}) as the number of direct references expected for each loaded cache line.  The metric does not consider cache reuse and only credits multiple elements from the same path in the same cache line.  The baseline breadth-first layout has an \textsf{EU} of 1 with the simplifying assumption that accessing each subsequent level of a tree guarantees a cache miss for all levels greater than two.  For \textsf{Stat} we calculate the \textsf{EU} as:$ EU_\textsf{Stat} = 1+bias(1+bias(1+bias)) $.  Where bias is defined as $ max(LC, RC)/PN $, where $LC$, $RC$, and $PN$ are the cardinalities of the left child node, right child node, and parent node respectively.  The 1 in $EU_\textsf{Stat}$ represents the guaranteed use of one element of the loaded cache line.  The remaining biases in the equation represent the likelihood of using each subsequent element in the cache line.  This assumes two nodes per cache line and use of an adjacent cacheline prefetching enabled CPU provides an additional two nodes per cacheline fetch.  Hence, a single loaded cache line could potentially be used four times.  \textsf{DF} layout doesn't consider node access frequency and so accessing either of the child nodes is equally likely, giving a bias of .5 and resulting in an (\textsf{EU}) of 1.85.  Because our findings indicate the vast majority of execution time is due to memory loads and thus far our optimizations don't modify the prediction algorithm, we used the following formula to determine an expected runtime of each of the optimizations:

\begin{multline} \label{eqn:cost}
 avg\_cache\_miss\_time = run\_time_{\textsf{BF}} / \\ (Forest_{avg\_depth})
 \end{multline}
 
\begin{multline}
 expected\_run\_time = avg\_cache\_miss\_time \\ * (Forest_{avg\_depth} - \#WuN)/EU_{Layout} 
\end{multline}
\\
Where \#WuN is an estimate of the number of well used nodes per prediction (e.g. interleaved nodes from lower levels and combined leaf nodes) which have a good chance of remaining in cache between prediction runs.  
This greatly oversimplifies reality in that it assumes that contiguous nodes produce cache hits and noncontiguous nodes result in cache misses.  This is not strictly true because noncontiguous nodes may still be resident in cache from prior use.  Contiguous nodes may also cross a page boundary which prevents use of hardware prefetching.  This model also ignores cache hits which, for large forests, should typically only occur for well used nodes such as low level nodes and the combined leaf nodes.

The next improvement interleaves the nodes of multiple trees into a single node array, called a bin, in order to encode parallel memory access.  The classification process concurrently navigates all trees in a bin.  This layout takes advantage of the fact that a parent node is accessed about twice as often as each of it's children and so nodes at lower levels of a tree are accessed far more often then nodes at higher levels.  By interleaving these lower level nodes from multiple trees contiguously in a bin, a single cache line fetch can load nodes from multiple trees.  The density of well used nodes at the beginning of each bin increases the likelihood of the hardware prefetcher loading useful nodes, thereby increasing the \textsf{EU}.  The \textsf{Bin} panel of Fig.~\ref{fig:layouts} shows the root (level 1) nodes interleaved in the layout and all higher level nodes are laid out statistically one tree at a time.  In addition to increasing the \textsf{EU}, this layout also reduces cache pollution, which increases the likelihood of well used nodes remaining in cache.  For this optimization to be effective, we also need to evaluate multiple trees in each bin concurrently (see Section~\ref{sec:exec}).

There are two parameters used when interleaving nodes. (1) The stripe width determines how many trees are binned into a single array.  Wider stripes encode more memory parallelism for each fetch at low levels in the tree.  Multiple stripes define independent arrays that can be distributed for parallel evaluation across cores or a cluster, so we need enough stripes to occupy parallel resources.  (2) Interleaved depth determines how many lower levels of the forest are interleaved before reverting to a per-tree statistical layout at higher levels.  Interleaving too many levels results in breadth-first like behavior, so there is a danger here that interleaving may degrade performance.

\begin{figure}[t!]
  \centerline{\includegraphics[width=\linewidth]{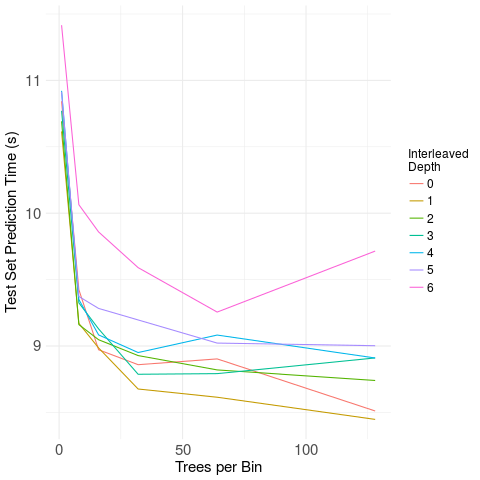}}
  \caption{Prediction time as a function of the number of trees interleaved into each bin and the depth of interleaving.  An interleaved depth of zero means all root nodes in the bin are contiguous in memory.  Interleaving higher level nodes degrades performance.}
  \label{fig:params}
\end{figure}

A design experiment helps choose these parameters and produces values that make sense in light of the properties of the processor memory hierarchy and forest characteristics.
Fig.~\ref{fig:params} explores the parameter space of depth and width for a trained forest.  The average depth traversed by observations in the test set is 16.86.  The figure measures average prediction time (lower is better) when classifying data using varying values of stripe width and interleaved depth.  Best performance occurs for 128 trees per bin and an interleaved depth of 1.  Each node is padded to $32$ bytes so that a cache line of $64$ bytes contains $2$ nodes.  Having several cache lines of data at depth 0 ensures hardware prefetching will assist with memory access.  
At depth 0, a contiguous range of memory is accessed and every node is used.
At depth 1, a contiguous range of memory is accessed and every other node in the range is used.  
At depth 2, one in every four nodes is used.  As we get deeper, there is less locality and thus benefits of binned interleaving diminish.  Optimal parameters vary by forest but in general larger bins perform better,  So, in a single core setting, using one large bin containing all trees may perform best, whereas in a multicore setting a bin size of $T/number of cores$ performs best by maximizing parallelism.

\begin{figure*}[tbhp]
  \begin{center} 
 \centerline{\includegraphics[width=\linewidth]{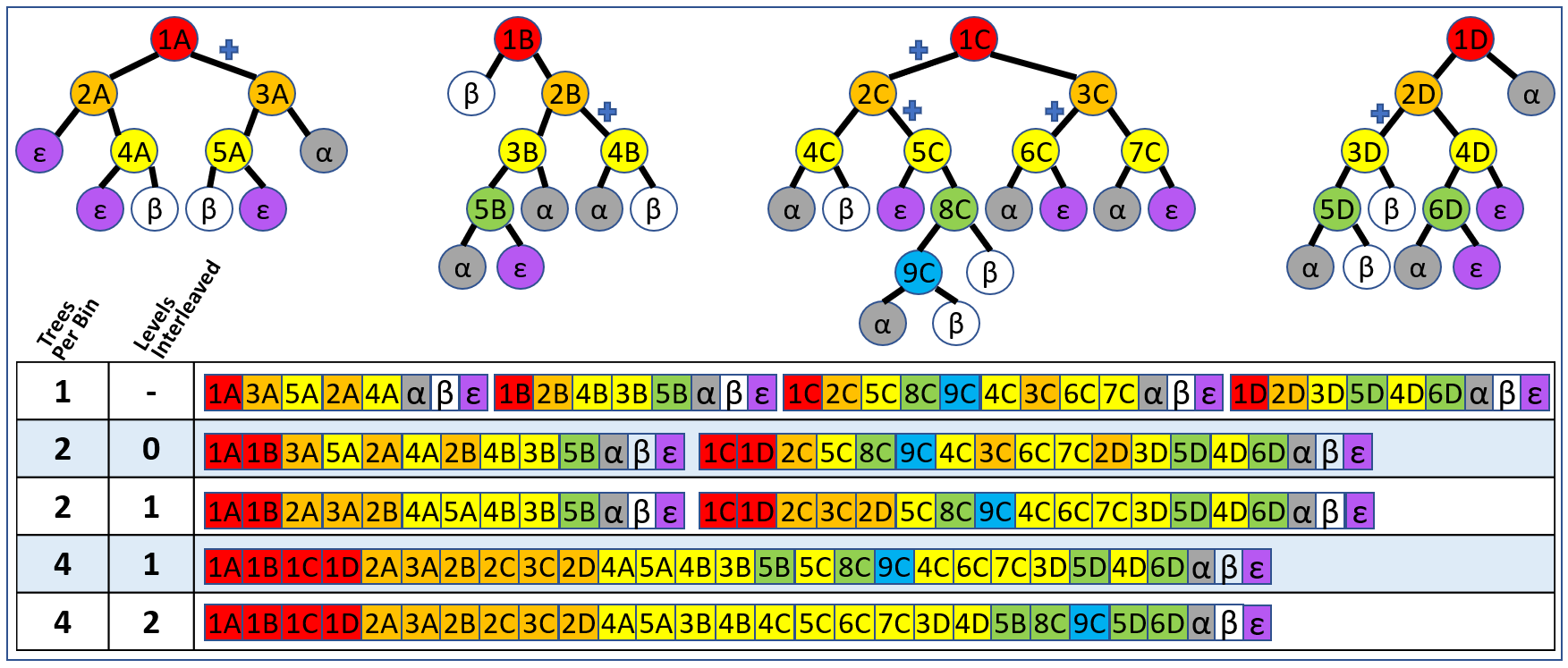}}
  \caption{Forest packing with the \textsf{Bin} memory layout.  We show different parameterizations of number of trees per bin and depth of interleaving.  Colors of internal nodes denote node level.} 
  \label{fig:fulllayout}
  \end{center} 
\end{figure*}

This layout combines binned interleaving for the top levels of a tree and statistical depth-first layout for the bottom levels in a tree.
Fig.~\ref{fig:fulllayout} demonstrates this for a forest of 4 trees and 3 classes with varying stripe widths and levels binned.

\subsection{Prefetching and Out-of-Order Execution}
\label{sec:exec}

In addition to memory parallelism, realizing performance gains requires fine-grained, low-level prefetching and scheduling.  
In \textsf{Bin+} we evaluate all trees in an interleaved bin together using the principle of overlapping execution and memory access.  Overlap and scheduling happens 
in the processor based on prefetching instructions that we emit and by defining a sufficient number of data independent tasks.  As we saw in Fig.~\ref{fig:params},
interleaving trees into bins provide only a modest gain of 20\% as a memory optimization.  However, binning makes overlapped execution possible,
which realizes more substantial performance gains.

We use simple policies to encode overlap, because more complex policies incur scheduling overheads that exceed gains.  Overlap happens among tens of memory accesses and hundreds to thousands of instructions.  Trying to control this fine-grained process in software is not practical.  
We assign a single thread to each interleaved bin that evaluates the trees in order.  For each node that we evaluate, we issue a prefetch instruction for the memory address of the chosen child node with the idea that useful work from other trees can be performed while the child node is being loaded into cache.  We proceed navigating through the levels of the tree in the same order that we processed the root.  This round-robin scheduling submits independent instructions for each tree.  In practice, the processor executes these independent instructions out-of-order as data is ready.  Round robin does not allow a fully asynchronous traversal of the tree in that all trees are within one level of each other at all times.  Yet, it is the only policy we identified that was lightweight enough to be effective.

\section{Evaluation}

We evaluate Forest Packing in order to quantify the benefits from memory layout and scheduling optimizations.  We start with an
overview experiment that demonstrates the performance improvement of Forest Packing versus unoptimized layouts.  Then, we breakdown the 
contributions of each individual optimization, applying them incrementally.  We explore parallelism in two scenarios.  First, we examine how to parallelize evaluation across a shared-memory multicore system to minimize the latency of a single classification.  Shared-memory scaling is slightly sublinear as cores compete and interfere in the memory system.  Forests are used this way for interactive applications.  We also study weak scaling across multiple nodes, corresponding to batch evaluation of data sets.  Distributed weak scaling shows the linear 
speedup that we expect from the data-independent evaluation of decision forests.

\subsection{Experimental Setup}

\begin{table}[tbhp]
  \begin{center}
    \caption{Dataset and Forest Information}
    \label{tab:forestInfo}
    \begin{tabular}{l l l l }
      \toprule 
      \textbf{} & \textbf{Allstate} & \textbf{Higgs} & \textbf{MNIST}\\
      \midrule 
      Training Observations & 500000 & 250000 & 60000\\
      Test Observations & 50000 & 25000 & 10000\\
      Number of Features in Dataset & 33 & 30 & 784\\
      Avg Number of Internal Nodes & 90020 & 23964.84 & 5008.93\\
      Number of Trees in Forest & 2048 & 2048 & 2048\\
      Avg Depth of Test & 25.48 & 20.81 & 16.86\\
      Largest Tree Depth in Forest & 74 & 65 & 50\\
      Average Bias & 50.000195 & 50.000755 & 50.00295\\
      \bottomrule 
    \end{tabular}
  \end{center}
\end{table}

We implement all of the optimizations described in Section~\ref{sec:methods} and apply them successively.  We start with the popular \rf{} implementation ranger \cite{ranger} in C++.  Even this implementation is 4x faster than the R-language implementation but it is slower than our breadth first implementation.  Thus, all comparative evaluations will be done against our re-implementation, \textsf{BF}.  We are unable to compare directly with XGBoost \cite{xgboost}, because the XGBoost implements a specific algorithm that creates very shallow trees.  We focus on standard decision forests that create deeper trees with a single class per leaf node.

All single core experiments were run on a 64-bit Ubuntu 16.04 platform with four Intel Xeon E7-4860V2 2.6 GHz processors, with 1TB RAM.  Multi node experiments were run on t2.xlarge AWS instances running Ubuntu 14.04, each with 4 vCPUs and 16 GiB RAM.  gcc 5.4.1 was used to compile the project using -fopenmp, -O3, and -ffast-math compiler flags.

We perform all experiments against three common machine-learning datasets that are widely used in competitions and benchmarks.
All datasets contain a combination of categorical, numeric, and/or scientific features.  MNIST is a popular numeric dataset of 60,000 handwritten digit images.  The \textsf{Higgs} dataset was used in a popular Kaggle competition  and contains 250,000 observations of both numeric and scientific features \cite{higgs}.  We use a subset of the Kaggle competition \textsf{Allstate} dataset which consists of categorical, numeric, and scientific features.\cite{allstate}  Further information about these datasets and resulting trained forests can be found in Table~\ref{tab:forestInfo}.

Each of the datasets contain both training and testing observations.  The training observations were used to train the forests and provide node traversal statistics when necessary.  The testing observations were only used for measuring prediction speeds and were not used to determine traversal statistics or resulting biases.

\subsection{Overall Performance}
\label{sec:overall}

\begin{figure}[tb]
  \centerline{\includegraphics[width=\linewidth]{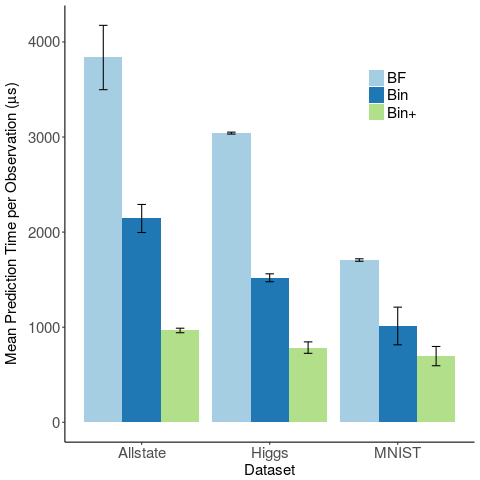}}
  \caption{Our optimized breadth first layout serves as a baseline for all performance gains.  \textsf{Bin} shows performance gains based on all layout optimizations.  \textsf{Bin+} shows additional performance gains from explicit prefetching and round-robin execution.  All experiments performed using a single core.}
  \label{fig:predict}
\end{figure}

Forest packing improves performance by a factor of four on the larger \textsf{Higgs} and \textsf{Allstate} datasets and by a factor of two for the
MNIST data.  Fig.~\ref{fig:predict} shows average prediction time per observation for three different algorithms on all datasets for our baseline \textsf{BF} layout and \textsf{Stat} and \text{Bin} layouts with a stripe width of 16 trees per bin and interleaved level of 3.  These parameters were chosen because they performed well for each of the tested forests, although not optimally, and so they provide consistency across experiments.  We were initially surprised that
memory layout alone did not improve performance more substantially.  The difference between \textsf{BF} and \textsf{Stat} is only 30-60\% on these benchmarks.  
This led us to look at scheduling more closely.  Adding scheduling realizes the full speedup.  The performance landscape is complex, 
because memory and scheduling interact.
The overlap between computation and memory access requires a large number of trees per interleaved bin so that the prefetch completes prior to round-robin
evaluation of the next level in each tree.  A large number of trees also gives the processor more independent work and more opportunities to reorder computation.

\subsection{Comparing Memory Performance with Estimates}

\begin{figure}[t]
  \centerline{\centerline{\includegraphics[width=\linewidth]{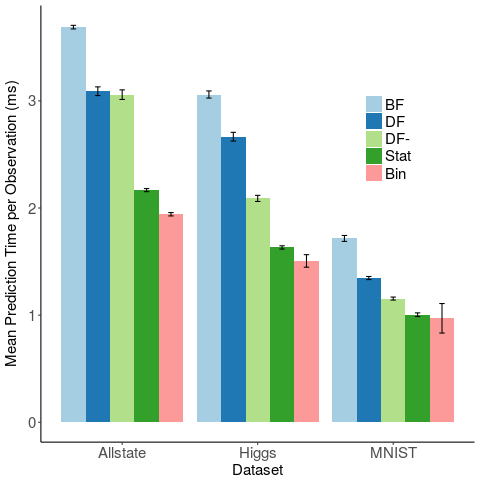}}}
  \caption{Effects of memory layout strategies on various datasets.  Dataset characteristics can be found in table Table~\ref{tab:forestInfo}.  All experiments performed using a single core.}
  \label{fig:versus}
\end{figure}

We now turn to a detailed study of the contribution of each of the layout optimizations and compare the experimental performance with the estimated performance.
Performance estimates use the simplified cache model described in Section~\ref{sec:MemLayout} and assume the classification workload is drawn from the same distribution used to train the forest.  Fig.~\ref{fig:versus} shows the time per observation when applying successive optimizations \textsf{DF, DF-, Stat, Bin} with 16 trees per bin and an interleaved depth of 3.  Results such as those of Fig.~\ref{fig:params} showed these parameters to be good, although not optimal, for each of the tested forests,  The performance shown involves no prefetching and no out-of-order execution.  The execution time of each layout is dominated by cache hits and misses. 

\begin{figure}[t]
  \centerline{\centerline{\includegraphics[width=\linewidth]{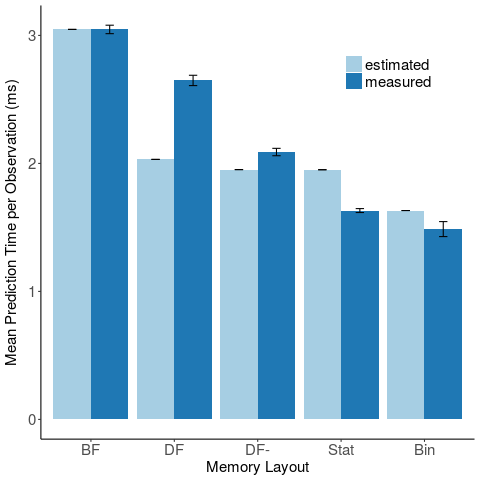}}}
  \caption{Estimated versus measured performance.  Performance estimations based on the dominating processing factor, memory loads, allows a pre-implementation assessment of a memory layout.  All experiments performed using a single core.}
  \label{fig:expected}
\end{figure}

\begin{figure*}[tbph]
 \begin{center}
\centerline{
  \subfigure[Latency]{\includegraphics[width=0.45\linewidth]{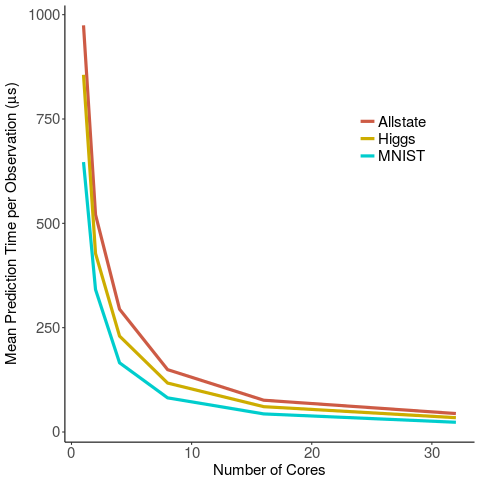}}
~
  \subfigure[Speedup]{\includegraphics[width=0.45\linewidth]{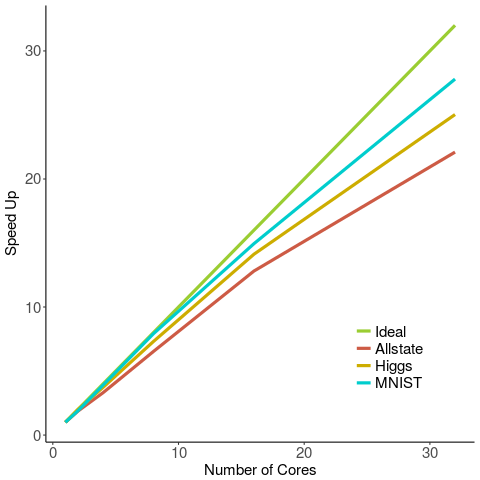}}
}
  \caption{Strong scaling on shared-memory multicores shows that memory interference results in sublinear scaling.}
  \label{fig:strong}
  \end{center}
\end{figure*}

Experimental performance tracks the \textsf{EU} based estimates, Fig.~\ref{fig:expected}.  We attribute the gap between estimate and measured time to several factors.  First, the simplified model relies on the average number of nodes traversed by the test set but not the distribution of leaf node depths.  This distribution could have a large effect on what could be considered a well used node.  To improve our model we would need to account for this distribution.  The second factor affecting the model is that all node accesses are considered to be either a hit or a miss with no regard to hit location.  In actuality the most used nodes will reside in lower level cache as compared to less frequently used nodes.  To take this into account would require estimating misses at each level of the cache hierarchy as well as memory page locality.  Another factor affecting estimated performance is the forest size.  The forest size reduction provided by (\textsf{DF-}) shows that forest size has an inverse relationship with performance.  By reducing forest size we improve locality and reduce cache pollution.  To improve the model would require calculating which nodes should be considered well used based on forest size.  The Stat layout performed better than expected based on the average bias for the forest.  This occurred because although the average bias of a forest is small, the bias of nodes, particularly low level nodes, could be high.  To better account for this phenomena we need a better overall measure of bias between nodes.  We also considered that our assumption about execution time remaining dominated by cache misses for each layout might be wrong.  But an almost identical runtime profile of the memory layouts showed an unchanged proportion of time waiting for data loads.

The trend for the forests used in these experiments are for larger training sets to have larger trees and require the average observation to travel to higher levels of a tree to make a prediction.  In general tree size and average observation prediction depth will be proportional to the training set size.  As such, it is expected that the larger forests will require higher training times.  In addition to requiring more memory accesses to make a prediction, large trees suffer from being more likely to overflow the cache and also have lengthier memory accesses.  At lower depths in the tree a cache miss may result in a memory access but it is also more likely that the required memory will be in the same active memory page.  As a prediction is required to traverse to higher depths in the tree it is more likely that subsequent memory accesses will also require changing the active memory page which requires more time than a read from an active memory page.  This further reveals why prefetching and round-robin execution is important, because it allows for several possibly long memory accesses to be processed simultaneously vs sequentially.

We were surprised to discover that an almost imperceptible difference in average bias between child nodes (Table~\ref{tab:forestInfo}) would provide an almost 30\% reduction in execution times when using the \textsf{Stat} layout.  At first we assumed this small bias was due to the characteristics of the datasets, but then we discovered two more plausible reasons.  First, \rf{}'s splitting function, gini, highly favors splits of equal cardinalities.  Second, the forests were trained to maximum depth which, for complex datasets, leads to single observation leaf nodes.  Since leaf nodes account for half of the nodes in a tree and most of the leaf nodes in the forests result from a perfect 50-50 split, it is reasonable to conclude that our bias calculation is dominated by perfect splits at these leaf nodes.  It is not uncommon for many real-world data sets to have large class imbalances which will lead to even larger biases.  In these cases, we see \textsf{Stat}, providing an even more significant savings.  The benchmark data we used does not have this property.

\subsection{Parallel Evaluation to Reduce Latency}
  
We study the effect of parallelizing classification in a shared memory multicore to reduce latency.  This is a strong scaling experiment that runs a single classification on one to 32 processors.  The bins are distributed equally among and bound to cores.  For all experiments, we use 16 bins per tree and interleave two levels.  Shared-memory parallelism reduces latency effectively, but scaling is sublinear because classification in decision forests is memory bound.  Fig.~\ref{fig:strong}(a) shows that
multicore parallelism drives per classification latency down from a prediction time of 600-1100 microseconds at one core to 20-50 microseconds at 32 cores for an overall speed-up factor of about 26.  Looking at the same data
in a speedup chart makes parallel inefficiencies more visible.  We realize an Amdahl number 
of between .987 and .997 for different datasets.  We expected this number to be 1.0 because all bins, and thus thread workloads, are independent.  We attribute this small deviation from perfect parallelism to the skewed nature of observation traversal depths.  The execution time of a thread is determined by the length of the longest path traversed through all trees in the thread's assigned bins.  And similarly, the execution time for an observation is determined by the thread with longest execution time.

\subsection{Weak Scaling in Distributed Memory}

Decision forests are trivially parallelizable and exhibit perfect speedup when distributed
across the multiple nodes of a cluster.  This has been shown in previous systems \cite{xgboost}
that have integrated decision forests with HPC schedulers.  Unlike shared-memory multicore, 
distributed systems scale the memory resource with the computation.  We use weak scaling
experiments to demonstrate that forest packing has this property.  This experiment corresponds 
to a throughput-oriented, batch classification workload that classifies a large number of data points at one time.  This distributed version of our system was not designed to reduce latency.  As implemented there is not enough work required for the classification of a single observation to overcome the startup costs of distributing work to a cluster.  This may not always be the case though.

\begin{figure}[tb]
  \centerline{\includegraphics[width=\linewidth]{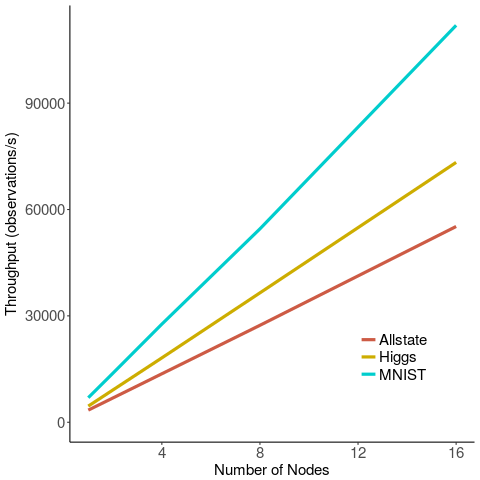}}
  \caption{Weak scaling of decision forest classification.  Throughput grows linearly when adding additional distributed processing nodes.}
  \label{fig:weak}
\end{figure}

Fig.~\ref{fig:weak} shows that throughput scales linearly as we increase the input size in proportion to the number of nodes in a cluster.
There are small fixed startup costs from distribution, but they do not accumulate as we progress to higher degrees of parallelism.  We use a different configuration of nodes for just this experiment to reduce costs.  A trained forest for each of the datasets was placed on an AWS node which was then cloned.  Weak scaling is demonstrated by starting cloned instances to increase compute capacity.  Observations are then processed by each of the running nodes which linearly increases observation processing throughput as expected.  

\section{Discussion and Conclusions}

We have addressed the poor memory performance of classification in decision forests with a series of scheduling and memory layout
optimizations that transform performance.  For random forests, performance increases by 50 times over the R-language implementation and 
by 4 times over a popular C++ implementation using a similar layout.  With improved performance, decision forests become more attractive for inclusion in HPC applications and workflows.  

We have not answered the question of how forest packing can be used alongside the shallow-tree optimizations of XGBoost \cite{xgboost}.
Our efforts have focused on common applications of decision forests that build deeper trees that partition data until there is a 
single class per leaf node.  XGBoost does not optimize across trees but all of the memory layout techniques of forest packing 
should still apply.  This remains to be implemented and verified.

We expect decision forests to be adopted increasingly in HPC and that the performance improvements outlined in this 
paper will ease that adoption.  Machine learning, mostly deep learning, has been demonstrated to be beneficial.
Machine learning can replace numerical models with compact, more accurate representations.  It can guide parameter search over
many simulations to find solutions faster and it can reduce data, identifying interesting features in massive outputs.
The adoption of deep learning has occurred despite its limitations: the results of deep learning cannot be interpreted in a human-understandable way and deep-learning is computationally intensive.  Decision forests are interpretable and efficient and will be preferred to deep learning for many applications.  Our work reduces the barrier to their adoption in supercomputing.


\bibliographystyle{unsrt}
\bibliography{main}

\end{document}